\documentclass[aps,twocolumn,showpacs,superscriptaddress,10pt]{revtex4-1}
\usepackage[utf8]{inputenc}
\usepackage{graphicx,amsmath,color}
\usepackage{amssymb}
\usepackage{amsthm}
\usepackage{times}
\usepackage{bbm}
\usepackage{bbold}
\usepackage[colorlinks,bookmarks=true,citecolor=blue,linkcolor=red,urlcolor=blue]{hyperref}

\newcommand{\je}[1]{\textcolor{black}{#1}}

\begin{document}

\title{Topology of density matrices}

\author{Jan Carl Budich}

\affiliation{Institute for Theoretical Physics, University of Innsbruck, 6020 Innsbruck, Austria}
\affiliation{Institute for Quantum Optics and Quantum Information, Austrian Academy of Sciences, 6020 Innsbruck, Austria}

\author{Sebastian Diehl}
\affiliation{Institute of Theoretical Physics, TU Dresden, D-01062 Dresden, Germany}

\date{\today}
\begin{abstract}
We investigate topological properties of density matrices motivated by the question to what extent phenomena such as topological insulators and superconductors can be generalized to mixed states in the framework of open quantum systems.
The notion of geometric phases has been extended from pure to mixed states by Uhlmann in [Rep. Math. Phys. {\bf{24}}, 229 (1986)], where an emergent gauge theory over the density matrices based on their pure-state representation in a larger Hilbert space has been reported. However, since the uniquely defined square root $\sqrt{\rho}$ of a density matrix $\rho$ provides a global gauge, this construction is always topologically trivial. Here, we study a more restrictive gauge structure which can be topologically non-trivial and is capable of resolving homotopically distinct mappings of density matrices subject to various spectral constraints. Remarkably, in this framework, topological invariants can be directly defined and calculated for mixed states. In the limit of pure states, the well known system of topological invariants for gapped band structures at zero temperature is reproduced.  We compare our construction with recent approaches to Chern insulators at finite temperature. 
\end{abstract}
\maketitle

\section{Introduction}
With the identification of geometric phases in quantum physics \cite{Berry,Simon,WilZee}, an emergent classical gauge structure in elementary quantum mechanics has been revealed. Several fundamental discoveries of contemporary quantum physics such as the integer quantum Hall effect \cite{Klitzing1980,Laughlin1981,TKNN1982,Kohmoto1985} and, more recently, the periodic table of topological insulators and superconductors \cite{Schnyder2008,Kitaev2009,Ryu2010,HasanKane,XLReview} can be theoretically described as topological properties of such emergent gauge theories in the framework of Bloch bands: The local $U(n)$ gauge degree of freedom there consists in the choice of an arbitrary orthonormal basis of Bloch states $\lvert u_k^\alpha\rangle$ that span the projection $P(k)=\sum_{\alpha=1}^{n}\lvert u_k^\alpha\rangle\langle u_k^\alpha\rvert$ onto the $n$ occupied bands of an insulating band structure at lattice momentum $k$. The covariant derivative
for such gauge theories has been constructed by Kato as early as 1950 in his proof of the adiabatic theorem of quantum mechanics \cite{Kato1950}. Topological invariants of this gauge structure distinguish homotopically distinct mappings $k\mapsto P(k)$, i.e., topologically inequivalent band structures. The consideration of generic symmetries \cite{Altland} refines this homotopy classification and leads to the periodic table of topological invariants \cite{Schnyder2008,Kitaev2009,Ryu2010,HasanKane,XLReview} (see Ref. \cite{KennedyZirnbauer} for a recent rigorous discussion).\\
 
Real physical systems, however, are not perfectly isolated from their environment and are to be described by a mixed state density matrix $\rho$ rather than a pure state wave function. It is hence natural to ask whether the concept of geometric phases in quantum physics can be generalized to the realm of density matrices and their quantum mechanical evolution. This question has been answered in the affirmative in a series of papers by Uhlmann \cite{Uhlmann1986,Uhlmann1991,Uhlmann1993}, who viewed the redundancy in the purification of density matrices, i.e., in the representation of $\rho$ as a pure state in a larger Hilbert space, as a gauge degree of freedom: $\rho=w w^\dag=(wU)(wU)^\dag$, where the Hilbert Schmidt operator $w$ denotes a purification of $\rho$ and $U$ is a unitary operator. In this context, the Uhlmann-connection defining parallel transport of Hilbert Schmidt operators along a path $t\mapsto \rho(t)$ of density matrices is given by the constraint \cite{Uhlmann1986}
\begin{align}
\dot w^\dag w-w^\dag \dot w=0,
\label{eqn:UhlmannCon}
\end{align}
where the dot denotes the derivative with respect to $t$.\\

The purpose of our present work is to unravel the topological properties of various gauge structures over the space of quantum mechanical density matrices. Since the purification $w=\sqrt{\rho}$, uniquely defined for every density matrix $\rho$, provides a global gauge, the general purification scheme in terms of Hilbert Schmidt operators as considered by Uhlmann is always topologically trivial. However, we would like to point out that this scheme can also be applied to density matrices of pure states which then would also yield a topologically trivial gauge theory -- despite the fact that pure states give rise to a rich spectrum of topological phenomena as mentioned above. Thus, the fact that some topologically trivial gauge structure can be constructed over the space of density matrices by no means implies that there are no topological features to be discovered. In fact, more generally speaking, every topological gauge theory can be embedded into a topologically trivial theory with a larger gauge degree of freedom over the same base space \cite{note2}. This raises the natural question whether a more restricted notion than general purification of density matrices in terms of Hilbert Schmidt operators could be employed to reveal topological aspects of families of density matrices $\rho(k)$ parameterized by a lattice momentum $k$.\\  

{{\emph{Key results --}}}
We investigate an ensemble of pure states (EOPS) scheme in terms of non-orthonormalized pure states $\lvert \tilde \psi^\alpha\rangle$ satisfying 
\begin{align}
\rho=\sum_\alpha \lvert \tilde \psi^\alpha\rangle\langle \tilde \psi^\alpha\rvert
\label{eqn:eopsproj} 
\end{align}
as a gauge structure over the space of density matrices $\rho$. This construction draws intuition from the entanglement theory of mixed states where, e.g., the entanglement of formation \cite{EntanglementFormation} is defined in terms of similar ensembles of pure states.
Starting from the spectral representation $\rho=\sum_\alpha p_\alpha \vert \psi_\alpha\rangle\langle \psi_\alpha\rvert$, a natural EOPS is $\lvert \tilde \psi_\alpha \rangle = \sqrt{p_\alpha} \lvert \psi_\alpha\rangle$. Under arbitrary unitary rotations of this frame, the pure states are no longer mutually orthogonal but still project down to the same density matrix $\rho$ via Eq. (\ref{eqn:eopsproj}).
The resulting gauge theory of non-orthogonal frames over the density matrices is a generalization of the pure state case.
For invertible density matrices without any further spectral assumptions, it is equivalent to Uhlmann's construction and thus topologically trivial. Interestingly, if assumptions on the spectral degeneracy of the density matrices are made, the present scheme can accommodate topologically non-trivial mixed states. Topologically inequivalent mappings $k\mapsto \rho(k)$ in this framework cannot be continuously deformed into each other without either violating the spectral assumptions or breaking possible protecting physical symmetries. Our construction formalizes the notion of a purity gap proposed in Refs. \cite{DiehlTopDiss,BardynTopDiss} -- a purity gap closing is a level-crossing in $\rho$ and thus a violation of the underlying spectral assumptions that makes a change in the topology possible. Under non-equilibrium conditions, a physical system may be non-ergodic and may thus be described by a singular density matrix of rank $n<N$, where $N$ is the dimension of the Hilbert space. Also in this case, the EOPS gauge structure can become topologically non-trivial. The topological invariants associated with the proposed gauge structure are constructed in terms of Uhlmann's connection (see Eq. (\ref{eqn:UhlmannCon})) without reference to the pure state case which is, however, correctly reproduced as a limiting case here.

Finally, we discuss notable differences to recent publications \cite{Delgado1D,Arovas2014, Delgado2014} where topological invariants for two-banded one-dimensional systems and Chern insulators \cite{QAH} at finite temperature, respectively, have been proposed that are not characteristic classes of a gauge theory. We illustrate with an explicit example that a two-dimensional system at finite temperature is in general not uniquely characterized by a single invariant of that kind.\\

{{\emph{Outline}--}}
This article is structured as follows. In Section \ref{sec:reviewres}, we review how a gauge structure emerges for both pure and mixed states in elementary quantum mechanics.  In Section \ref{sec:mixtop}, topological invariants for density matrices with various spectral assumptions are defined. We compare the present approach to the construction of Uhlmann phase winding numbers that have recently been proposed \cite{Arovas2014, Delgado2014}  to classify thermal Chern insulators in Section \ref{sec:comments}. Concluding remarks are presented in Section \ref{sec:conclusion}.

\section{Emergent gauge structures from pure to mixed states}
\label{sec:reviewres}
\subsection{Adiabatic time evolution of pure states}
\label{sec:pureintro}
A natural and conceptually simple scenario for the occurrence of geometric phases is the time evolution of a quantum mechanical system with a Hamiltonian $H(R(t))$ that depends adiabatically on time via some control parameters $R(t)$ \cite{note1}. As a side remark, we note that the concept of geometric phases has been generalized to non-adiabatic \cite{AAphase} and non-cyclic \cite{Pphase,PphasePRL} evolution. The non-degenerate ground state of $H(R)$ is denoted by $\lvert R\rangle$. 
Since the system cannot leave its instantaneous ground state when initially prepared in $\lvert {R(t_0)}\rangle$, the geometric constraint 
\begin{align}
P(t)\lvert \Psi(t)\rangle=\lvert \Psi(t)\rangle
\label{eqn:pureconstraint} 
\end{align}
with $P(t)= \lvert{R(t)}\rangle\langle {R(t)}\rvert$ the projection onto the ground state, is imposed on the solution $\lvert \Psi(t)\rangle$ of the time dependent Schr\"odinger equation in the adiabatic limit.
This constraint, reflecting the absence of any dynamical level transitions, implies the form $\lvert \Psi(t)\rangle=\text{e}^{i(\phi(t)-\phi_D(t))}\lvert {R(t)}\rangle$, where $\lvert {R(t)}\rangle$ is a family of ground states with a smooth relative phase, i.e., a gauge of ground states. $\phi_D(t)=\int_{t_0}^t \text{d} \tau E_0(\tau)$ with the ground state energy $E_0$ is called the dynamical phase. The additional phase factor $\text{e}^{i\phi(t)}$ of the state vector $\lvert\psi(t)\rangle=\text{e}^{i\phi_D(t)}\lvert \Psi(t)\rangle$ relative to the gauge $\lvert R(t)\rangle$ reveals a deep geometric ``principle of least effort" that nature employs in the adiabatic limit of quantum mechanics: $\lvert\psi(t)\rangle$  is the shortest path in Hilbert space that satisfies Eq. (\ref{eqn:pureconstraint}), shortest as measured by $L=\int_{t_0}^t\text{d}\,\tau \sqrt{\langle \dot \psi(\tau) \vert \dot \psi(\tau)\rangle}$ with ``velocity''  $\vert \dot \psi(\tau)\rangle$, i.e., by the metric induced by the inner product in Hilbert space. To systematically construct this path, it is helpful to decompose its tangent vector into two orthogonal components as $\lvert \dot\psi(t)\rangle=P(t)\lvert \dot\psi(t)\rangle+(1-P(t))\lvert \dot\psi(t)\rangle$. The first component is called the vertical part $\lvert \dot \psi \rangle_V$ and generates an evolution that stays within the projection $P(t)$, i.e., a mere change of phase of $\lvert \psi(t)\rangle$. The second part in contrast generates an evolution perpendicular to $P(t)$ in the sense of the inner product and is consequently called the horizontal part $\lvert \dot \psi \rangle_H$. From $\lVert \dot \psi\rVert^2=\lVert\dot \psi_V\rVert^2+\lVert \dot \psi_H\rVert^2$, it is easy to see that the tangent vector to the shortest path must be purely horizontal, i.e., 
\begin{align}
P(t)\lvert \dot\psi(t)\rangle=0,   
\label{eqn:parallelcond}
\end{align}
or, equivalently for non-degenerate ground states, $\langle \psi(t)\vert \dot \psi(t)\rangle=0$. This parallel-transport prescription immediately determines
\begin{align}
\phi(t)=i\int_{t_0}^t \text{d}\,\tau \langle{R(\tau)}\rvert \frac{d}{d\tau}\lvert{R(\tau)}\rangle.
\label{eqn:BerryPhase}
\end{align}
The quantity $\mathcal A_{\frac{d}{dt}}(R(t))=\langle{R(t)}\rvert \frac{d}{dt}\lvert{R(t)}\rangle$ transforms under a gauge transformation $\lvert R\rangle\rightarrow \text{e}^{i\chi(R)}\lvert R\rangle$ as $\mathcal A_{\frac{d}{dt}}\rightarrow \mathcal A_{\frac{d}{dt}}+i\frac{d}{dt}\chi(R(t))$, i.e., as a $U(1)$ gauge field. The geometric phase $\phi^\gamma=i\int_\gamma \mathcal A ~(\text{mod } 2\pi)$ associated with a loop $\gamma$ in parameter space is gauge invariant. 
This construction can be immediately generalized to $n$-fold degenerate ground states where the projection  $P(t)$ in Eq. (\ref{eqn:parallelcond}) becomes
\begin{align}\label{eqn:proj}
P(t)=f(t) f^\dag(t) = \sum_{j=1}^n \lvert {R^j(t)}\rangle \langle {R^j(t)}\rvert
\end{align}
with the ground state manifold basis or frame $f = (\lvert {R^1}\rangle, ... , \lvert {R^n}\rangle)$. We note the independence of the projection under $U(n)$ basis transformations $f \to f U$. Thus, instead of a mere phase factor, the local gauge degree of freedom is then a $U(n)$ basis transformation on the ground state manifold. In this case, the geometric phase or, in mathematical terms, the $U(1)$ holonomy ${e}^{i\phi^\gamma}$ is replaced by the $U(n)$ holonomy
\begin{align}
U^\gamma=\mathcal T \text{e}^{-\int_\gamma \mathcal A},
\label{eqn:WilZee}
\end{align}
 where $\mathcal A^{jl}=\langle R^j\rvert d\lvert R^l\rangle$ is the non-Abelian gauge field or connection and $\mathcal T$ is the time ordering operator along the cyclic path $\gamma$. 

We finally note that the connection can be expressed in a manifestly gauge invariant or basis independent form as $\mathcal A_K=-[(dP),P]$ \cite{Kato1950}. In this case, the geometric phase in the ground state manifold for a loop $\gamma$ starting at time $t_0$ is described by the propagator of the Schr\"odinger equation in the adiabatic limit, or the holonomy operator
\begin{align}
U^\gamma_K= P(t_0)\mathcal T \text{e}^{-\int_\gamma \mathcal A_K}P(t_0).
\label{eqn:Kato}
\end{align}
$U^\gamma$ as occurring in Eq. (\ref{eqn:WilZee}) is then just the gauge (basis) dependent representation matrix of $U^\gamma_K$.
\\
\subsection{From geometric phases to topological band structures}
\label{sec:gtot}
Let us consider an insulating band structure in $d$ spatial dimensions with $n$ occupied Bloch states $\lvert u_k^\alpha\rangle$ below the Fermi energy that span the projection $P(k)=\sum_{\alpha=1}^{n}\lvert u_k^\alpha\rangle\langle u_k^\alpha\rvert$ at lattice momentum $k$. If we identify the energy gap relevant for the adiabatic approximation with the band gap of the insulator and the parameter manifold with the Brillouin zone (BZ) in which the lattice momentum is defined, a gauge structure can be defined in complete analogy to Section \ref{sec:pureintro}. The projection $P(k)$ defines an $n$-plane in the $N$-dimensional Hilbert space defined at every momentum $k$ in the BZ, where $N$ is the total number of bands (occupied and empty) of the model system. $P(k)$ is thus by definition a point on the Grassmann manifold $G_n(\mathbb C^N)=U(N)/(U(n)\times U(N-n))$. Topological invariants of this gauge structure then distinguish topologically inequivalent mappings $k\mapsto P(k)$, that is homotopically inequivalent mappings from the BZ torus $T^d$ to $G_n(\mathbb C^N)$. For the case of particle number conserving insulators without additional symmetries outlined here, an integer invariant named the $m$-th Chern number \cite{Chern1946} is defined for even spatial dimensions $d=2m$ while all insulators are equivalent in odd spatial dimensions. Physical symmetries imply constraints on the form of the mapping $k\mapsto P(k)$. For example the anti-unitary time reversal symmetry $T$ yields $P(k)=TP(-k)T^{-1}$. Mappings (band structures) that would be topologically equivalent when breaking the relevant symmetries may be topologically inequivalent when maintaining the symmetries. This phenomenon is commonly referred to as symmetry protection of a topological state. Taking also into account superconducting band structures as well as all symmetries considered by Altland and Zirnbauer \cite{Altland} significantly refines the system of topological invariants and leads to a pattern which has been coined the periodic table of topological insulators and superconductors \cite{Schnyder2008,Kitaev2009,Ryu2010,HasanKane,XLReview}. In this work, we would like to address the question to what extent such invariants can be generalized to lattice translation-invariant quantum many body systems in a mixed state. While the relevant physical symmetries can readily be generalized to the realm of density matrices, the definition of topological invariants for mixed states from a gauge structure over the space of density matrices has not been discussed so far.  
However, a topologically trivial gauge structure for density matrices has been discovered by Uhlmann  \cite{Uhlmann1986,Uhlmann1991,Uhlmann1993} as we will review now.
Later on (see Section \ref{sec:speccon}), a more restrictive gauge structure that can be topologically non-trivial will be derived from Uhlmann's general construction.

\subsection{Parallel transport and geometric phases for density matrices}
\label{sec:Uhlmann}
A density matrix $\rho$ is a positive semi-definite operator with unit trace on a Hilbert space $\mathcal H$. Here, we consider $\text{dim} \mathcal H=N<\infty$. For the application to gapped band structures that we have in mind here, $N$ plays the role of the total number of bands of a model system. $\rho$ can be represented as a pure state $\lvert \psi\rangle$ on an extended Hilbert space $\mathcal H_A\otimes \mathcal H_B$ with $\mathcal H_A \simeq \mathcal H_B \simeq \mathcal H$ such that $\text{Tr}\left[O\otimes\mathbb{1}\lvert \psi\rangle\langle \psi\rvert\right]=\text{Tr}_A\left[O\rho\right]$ for any operator $O$ acting in $\mathcal H$, a prescription referred to as state purification. Equivalently, the purification can be represented in terms of a Hilbert Schmidt operator, i.e., a $N\times N$ matrix $w$ that satisfies
\begin{align}
\rho =\text{Tr}_B\left[\lvert \psi\rangle\langle \psi\rvert\right]  = w w^\dag,
\label{eqn:rhoproj}
\end{align}  
where the trace over the auxiliary Hilbert space $\mathcal H_B$ becomes a matrix multiplication. This description contains a redundancy since under $w\rightarrow w U$ with $U\in U(N)$, $\rho\rightarrow wUU^\dag w^\dag=\rho$ is unchanged. We note a formal analogy to the basis independence of the projection \eqref{eqn:proj}. Furthermore, the inner product $(w,v)=\text{Tr}[w^\dag v]$ for matrices again (cf. Section \ref{sec:pureintro}) defines a natural means to measure the length $L=\int_{t_0}^{t}\text{d}\,\tau \sqrt{(\dot w,\dot w)}$ of a path $t\mapsto w(t)$ of Hilbert Schmidt operators that purify a path $t\mapsto \rho(t)$ in the sense of Eq. (\ref{eqn:rhoproj}). The shortest possible path can again be obtained by decomposing the tangent vector $\dot w$ into a vertical part $\dot w_V$ and a horizontal part $\dot w_H$ with the help of the inner product. $\dot w$ is purely horizontal for the shortest path. Vertical vectors can be defined in terms of Eq. (\ref{eqn:rhoproj}) as tangent vectors to curves $s\mapsto w(s)$ that project to the same $\rho=ww^\dag$ for all $s$. Vertical vectors are hence of the form $\dot w_V=\left.\frac{d}{ds}wU(s)\right|_{s=0}=wu$ with $U(s)=\text{e}^{us}\in U(N)$. 
Horizontal vectors $\dot w_H$ are defined as orthogonal to all vertical vectors, i.e., $\text{Tr}[\dot w_H^\dag wu]=0$ for all $U(N)$-generators $u=-u^\dag$. 
Adding the hermitian conjugate to this condition yields $\text{Tr}[u(\dot w_H^\dag w-w^\dag \dot w_H)]=0$. 
This can be true for every antihermitian $u$ only if the second antihermitian factor is zero by itself, i.e., the tangent vector $\dot w$ is horizontal if \cite{Uhlmann1986,Uhlmann1991,Uhlmann1993},
\begin{align*}
\dot w^\dag w-w^\dag \dot w=0.
\end{align*}
Eq. (\ref{eqn:UhlmannCon}), repeated here for convenience, is the density matrix analog of Eq. (\ref{eqn:parallelcond}).

Let us now construct the $U(N)$-gauge field associated with this parallel transport prescription, in a form that is amenable to practical calculations. For a given $\rho=\sum_j p_j \lvert j\rangle\langle j\rvert $, the uniquely defined square root $\sqrt{\rho}=\sum_j \sqrt{p_j}\lvert j\rangle\langle j\rvert$ defines a generic purification $w=\sqrt{\rho}$. In fact every purification can by means of a polar decomposition be written as 
\begin{align}\label{eqn:wrhoU}
w=\sqrt{\rho} U \text{ with } U\in U(N)
\end{align} 
in some analogy to the polar decomposition of a complex number. Let us consider a loop $\gamma:t\mapsto \rho(t),~t\in[0,T]$ in the space of density matrices.
We denote by $t\mapsto w(t)=\sqrt{\rho(t)}U(t)$ the parallel-transport of an arbitrary initial purification $w(0)=\sqrt{\rho(0)}U(0)$ along this loop $\gamma$ and call the geometric phase $H_U^\gamma=U(T)U(0)^\dag$ its Uhlmann holonomy. With $t_i=i T/M,~ M\in \mathbb N$, we can express $H_U^\gamma$ as
\begin{align}
H_U^\gamma=\lim_{M\rightarrow \infty}\prod_{i=1}^M U(t_i)U^\dag(t_{i-1}),
\end{align}
where the product is ordered from right to left with increasing $i$. We now wish to obtain the analog of Eq. \eqref{eqn:WilZee} in an explicit form. To this end, we first obtain the parallel-transported $U(t)$ in Eq. \eqref{eqn:wrhoU} at infinitesimally neighbouring points in time. Eq. (\ref{eqn:UhlmannCon}) can be shown to be equivalent to
\begin{align}
w^\dag(t+\epsilon) w(t)=w^\dag(t)w(t+\epsilon) \ge 0
\label{eqn:posparallel}
\end{align}
to leading order in $\epsilon$; in particular, the hermitian matrix $w^\dag(t+\epsilon) w(t)$ is positive semi-definite. Defining $V=w^\dag(t+\epsilon)w(t)=U^\dag(t+\epsilon)\sqrt{\rho(t+\epsilon)}\sqrt{\rho(t)}U(t)\ge 0$ we immediately verify
\begin{align*}
\sqrt{\rho(t+\epsilon)}\sqrt{\rho(t)}=H U(t+\epsilon)U^\dag(t)
\end{align*}
 with the semi-positive hermitian matrix $H=U(t+\epsilon)V(t)U^\dag(t+\epsilon)$. Hence, the singular value decomposition
 \begin{align*}
 \sqrt{\rho(t+\epsilon)}\sqrt{\rho(t)}=L DR^\dag
 \end{align*}
with unitary $L,R$ yields $H=L D L^\dag$ and, more importantly \cite{Arovas2014},
\begin{align}
U(t+\epsilon)U^\dag(t)=L R^\dag.
\label{eqn:connectionsvd}
\end{align}
With this recipe the connection $\mathcal A_U=-\dot U(t) U^\dag(t)=\lim_{\epsilon\rightarrow 0}\frac{1}{\epsilon}(1-U(t+\epsilon)U^\dag(t))$ can be computed explicitly. The desired holonomy then reads as
\begin{align}
H_U^\gamma = \mathcal T \text{e}^{-\int_\gamma \mathcal A_U}.
\label{eqn:Uhlmannhol}
\end{align}
We note that $\mathcal A_U$ is invariant under a rescaling $\rho(t)\rightarrow \lambda(t) \rho(t)$ with a strictly positive real function $\lambda(t)>0$, which affects the singular values $D$ alone. Therefore, the normalization of the density matrix is not relevant for geometrical and topological considerations.

\section{Topological aspects of mixed states}
\label{sec:mixtop}

In order to discuss topological features of emergent gauge theories for mixed states, it is natural to view the above decomposition of tangent vectors into vertical and horizontal components as a connection on a principle fiber bundle (PFB) (see e.g., Ref. \cite{Nakahara}). 

It is well known that the PFB of general Hilbert-Schmidt operators is topologically trivial since $\rho\mapsto\sqrt{\rho}$ is a global section (gauge). At first sight, this may seem discouraging. On the other hand, even the special case of pure states can be viewed as a topologically trivial gauge structure in terms of Hilbert Schmidt operators. In contrast, it is well known that the gauge structure for pure states discussed in Section \ref{sec:pureintro} can very well be topologically non-trivial, phenomena such as topological insulators being striking ramifications of this possibility. More concretely, the frame bundle of orthonormal frames $f_R$ spanning the projection $P(R)$ can be topologically non-trivial. However, it can always be viewed as a sub-bundle of the topologically trivial Hilbert-Schmidt bundle of arbitrary quadratic matrices $w(R)$ that satisfy $P(R)=w(R)w^\dag(R)$. In other words, the level of description determines whether topological aspects of pure states can be resolved.
More generally speaking, every gauge theory (PFB) can be viewed as a sub-bundle of a topologically trivial gauge theory (PFB) with a bigger gauge degree of freedom but over the same base manifold. 
As for mixed state density matrices key questions are thus whether topologically non-trivial features exist and, if so, to identify a gauge structure which is able to reveal their topological content. Here we consider several physically motivated constraints under which a topologically non-trivial PFB of EOPS will be defined. 

\subsection{Triviality and structure of the Hilbert-Schmidt bundle}
Let us first review the case of arbitrary invertible $N \times N$ density matrices $\rho$ as considered by Uhlmann \cite{Uhlmann1986}, i.e., strictly positive matrices without any further constraints such as normalization of the trace. The space $M_N(N)$ of all Hilbert Schmidt operators $w$ purifying such density matrices via Eq. (\ref{eqn:rhoproj}) is then simply given by the group of all invertible matrices $GL(\mathbb C,N)$. This is because $\rho=ww^\dag$ is strictly positive and has the non-vanishing determinant $\lvert \text{det} (w) \rvert^2$ if and only if  $w\in GL(\mathbb C,N)$. $M_N(N)=GL(\mathbb C,N)$ is the total space of the Hilbert-Schmidt bundle which projects via $\Pi: w\mapsto ww^\dag$ onto the base manifold of strictly positive matrices $D_N(N)=GL(\mathbb C,N)/U(N)$. The quotient form of $D_N(N)$ is rooted in the fact that any regular matrix $M$ can be uniquely written as $M= \sqrt{MM^\dag}U, ~U\in U(N)$. For the same reason, $w$ with $ww^\dag=\rho$, can be uniquely represented as $w=\sqrt{\rho}U,~U\in U(N)$.
 Hence the fiber over $\rho$ is given by $G_\rho=\left\{\sqrt{\rho}U:U\in U(N)\right\}$, which is manifestly isomorphic to $U(N)$. $GL(\mathbb C,N)\overset{\Pi}{\rightarrow}D_N(N)$ thus defines a PFB. Due to the existence of the global gauge or section $\rho\rightarrow \sqrt{\rho}$, the PFB is topologically trivial, i.e.,
\begin{align}
GL(\mathbb C,N)=GL(\mathbb C,N)/U(N)\times U(N).
\label{eqn:trivialUhlmann}
\end{align}
\\
Now we consider singular density matrices, i.e., semi-positive $N\times N$ matrices with rank $n<N$. A naive analog of the regular case with fibers $G_\rho=\left\{\sqrt{\rho}U:U\in U(N)\right\}$ where $\rho$ now has rank $n$ does not give a PFB. The reason is that the right-action $w=\sqrt{\rho}U$ of $U(N)$ is no longer free, i.e., many $U\in U(N)$ give the same $w=\sqrt{\rho}U$ due to the $N-n$-dimensional null-space of $\sqrt{\rho}$. Simply restricting the $U$ to $U(n)$ acting on the support of $\sqrt{\rho}$ does not resolve this issue since this action would not be transitive, i.e., unable to reach all $w$ with $w w^\dag=\rho$. In Ref. \cite{DittmannRudolph1992}, it has been shown that $M_N(n)\overset{\Pi}{\rightarrow}D_N(n),~n<N$ with the same projection $\Pi: w\mapsto ww^\dag$ still defines a PFB. However, by the same argument as before, namely by the existence of the global section $\rho \mapsto \sqrt{\rho}$, also these bundles are topologically trivial.

\subsection{Ensemble of pure states gauge structure}
Instead of considering general Hilbert Schmidt operators, we define a EOPS bundle in terms of $n$ non-orthonormalized pure states $\lvert \tilde \psi^\alpha\rangle$ that project onto a given $N\times N$ density matrix $\rho$ of rank $n$ as (Eq. (\ref{eqn:eopsproj}) is repeated for convenience)
\begin{align*}
\rho=\sum_{\alpha=1}^{n} \lvert \tilde \psi^\alpha\rangle\langle \tilde \psi^\alpha\rvert 
\end{align*}
as a gauge structure over the space of density matrices $\rho$. Instead of viewing $\rho$ as a single pure state in a larger Hilbert space, EOPS in this scheme are plausible ensembles of pure states in the system Hilbert space.
A generic choice $\lvert \tilde \psi_\alpha \rangle = \sqrt{p_\alpha} \lvert \psi_\alpha\rangle$ along these lines is obtained from the spectral representation $\rho=\sum_\alpha p_\alpha \vert \psi_\alpha\rangle\langle \psi_\alpha\rvert$. Under arbitrary unitary rotations
\begin{align}
\lvert\tilde \varphi_\alpha\rangle= \lvert \tilde\psi_{\beta}\rangle U_{\beta\alpha},
\label{eqn:ensemblegauge}
\end{align}
the density matrix $\rho$ obtained from the projection (\ref{eqn:eopsproj})  is unchanged. However, the pure states $\lvert \tilde \varphi_\alpha\rangle$ are no longer mutually orthogonal. The matrix representation
\begin{align}
w=\left(\lvert \tilde \psi_1\rangle,\ldots,\lvert \tilde \psi_n\rangle,0,\ldots,0\right)
\label{eqn:DittmannEmbedding}
\end{align}
with $N-n$ zero columns defines a natural embedding of the space $P_N(n)$ of non-orthonormal $n$-frames into the space $M_N(n)$ of Hilbert Schmidt operators of rank $n$ \cite{DittmannRudolph1992}. $U$ as occurring in Eq. (\ref{eqn:ensemblegauge}) defines the local $U(n)$ gauge degree of freedom  of the PFB $P_N(n)\overset{\Pi}{\rightarrow}D_N(n)$. For regular  density matrices ($n=N$), it is easy to see that the space of all EOPS is identical to the space of all Hilbert Schmidt operators satisfying $\rho=ww^\dag$, i.e., $P_N(N)=M_N(N)$. Thus, for invertible density matrices with no further spectral constraints, the EOPS scheme is equivalent to the purification by pure state representation in a larger Hilbert space. From this observation, we immediately conclude that the EOPS bundle is topologically trivial for unconstrained regular density matrices. In the following, we will consider several constraints under which the EOPS bundle can become topologically non-trivial and explicitly construct the corresponding topological invariants.

\subsection{Gauge structure of spectrally constrained density matrices}
\label{sec:speccon}
{\emph{Non-degenerate density matrices --}}
The situation becomes more interesting if we impose certain conditions on the spectrum of the density matrix.
Let us first consider a regular {\emph{non-degenerate}} density matrix $\rho=\sum_\alpha p_\alpha \lvert \psi_\alpha\rangle \langle \psi_\alpha\rvert$, where we can now without loss of generality assume $p_1>p_2>\ldots>p_N$. 
For the hermitian operator $\rho$ with non-denerate eigenvalues, the projections $P_\alpha =\lvert \psi_\alpha\rangle \langle \psi_\alpha\rvert$ are mutually orthogonal and it is natural to order the columns $\sqrt{p_\alpha} \lvert\psi_\alpha\rangle$ of a Hilbert Schmidt representation (\ref{eqn:DittmannEmbedding}) with descending size of the spectral weights $p_\alpha$. 
This ordering will only be maintained under a subgroup of gauge transformations (\ref{eqn:ensemblegauge}) that are a direct sum of $N$  $U(1)$-transformations $\text{e}^{i\phi_\alpha}$ acting on the rays of eigenstates associated with the eigenvalues $p_\alpha$. Eq. (\ref{eqn:ensemblegauge}) then simplifies to
\begin{align}
\lvert \tilde \varphi_\alpha\rangle=\text{e}^{i\phi_\alpha}\lvert \tilde\psi_\alpha\rangle,\quad\alpha=1,\ldots,N.
\end{align}
The EOPS bundle thus constrained consists of $N$ $U(1)$-bundles $P_N(1)\overset{\Pi}{\rightarrow}D_N(1)$ which are subject to the constraint
\begin{align}
\sum_{\alpha=1}^{N}\lvert  \psi_\alpha\rangle\langle \psi_\alpha\rvert=\mathbb 1.
\label{eqn:sumconstraint}
\end{align}
While the individual $U(1)$-bundles can be topologically non-trivial, Eq. (\ref{eqn:sumconstraint}) enforces a zero sum rule for their topological invariants. The reason for this is analogous to the pure state case and can be intuitively understood as follows. The individual subspaces may exhibit a topologically non-trivial winding as a function of the control parameters (e.g. lattice momentum) in the total Hilbert space. This total space serves as fixed reference and thus by definition does not exhibit any "net-winding". Hence, the parameter dependences of the individual subspaces have to compensate each other in order to span the resolution of the identity (\ref{eqn:sumconstraint}) of the total embedding space at every point in parameter space.

The spectral constraint forbidding any level degeneracy can be somewhat relaxed by assuming that $m$ subsets of cardinality $n_1,\ldots,n_m$ with $N=\sum_{j=1}^{m}n_j$ of levels are allowed to be degenerate, but still have distinct eigenvalues from levels outside of their subset. Such constraints simply result in a restricted gauge group of the form $U(n_1)\times U(n_2)\times \ldots \times U(n_m)$. Again, the individual $U(n_j)$-structures may be topologically non-trivial but obey a zero sum rule due to Eq. (\ref{eqn:sumconstraint}).

{\emph{Purity gaps --}}
The notion of a purity gap \cite{DiehlTopDiss, BardynTopDiss} can be rationalized in this framework. To this end, we consider a spectral constraint along the lines of the above discussion where the $n$ largest eigenvalues of $\rho$ belong to one subset but are by assumption not degenerate with the $(n+1)$-th largest eigenvalue of $\rho$. Under these circumstances we can uniquely define a pure state projection $P=\sum_{\alpha=1}^{n}\lvert \psi_\alpha\rangle\langle \psi_\alpha\rvert$ which is closest to the mixed state defined by $\rho$. In Refs. \cite{DiehlTopDiss, BardynTopDiss}, the topological invariant of $\rho$ has been defined as the invariant of the corresponding pure state projection $P$. Below, we directly define topological invariants for mixed states in terms of the Uhlmann connection (\ref{eqn:posparallel}) restricted to the subspace associated with the $n$ largest eigenvalues of $\rho$. Their value is equal to the one obtained by performing an adiabatic deformation into a pure state along the lines of Refs. \cite{DiehlTopDiss, BardynTopDiss} but, quite remarkably, no reference to pure states is required here. A purity gap closing means that the $n$-th largest eigenvalue and the $(n+1)$-th largest eigenvalue of $\rho$ become degenerate, a singular-point at which the spectral constraint is violated and the topological invariant associated with the subspace of the $n$ largest eigenvalues is not well defined. As a consequence, this subspace can exchange topological charge with the eigenspace associated to the $(n+1)$-th largest eigenvalue and the topological invariant for the subspace of the $n$ largest eigenvalues may have changed when the purity gap reopens -- a topological phase transition by purity gap closing.\\

{\emph{Singular density matrices --}}
 A physical system under non-equilibrium conditions may not be ergodic and can thus be described by a singular density matrix, i.e., a density matrix of rank $n<N$. This constraint can be viewed as special case of the above discussion of spectral constraints: A subset of $n$ eigenvalues $p_\alpha$ is non-zero and hence not degenerate with the complement of $N-n$ non-zero eigenvalues. The $U(n)$ EOPS bundle associated with the non-zero subspace can be topologically non-trivial.

\subsection{Topological invariants}
The gauge structure of spectrally constrained density matrices defined in Section \ref{sec:speccon} allows for the definition of topological invariants directly in terms of the Uhlmann connection (see Section \ref{sec:Uhlmann}), i.e., without reference to pure states. To this end, we consider a subset of $n<N$ eigenvalues (eigenvectors) labeled by $j=1,\ldots,n$ that are by assumption non-degenerate with (orthogonal to) the $N-n$ remaining eigenvalues (eigenvectors) of $\rho$. We denote by $\mathcal P=\sum_{j=1}^n \lvert \psi_j\rangle\langle \psi_j \rvert$ the projection onto the rank $n$ density matrix $\hat \rho=\sum_j \lvert \tilde \psi_j\rangle\langle \tilde \psi_j \rvert$, i.e., $\hat \rho=\mathcal P \rho \mathcal P$. Eq. (\ref{eqn:DittmannEmbedding}) shows how the corresponding $U(n)$ EOPS bundle can be embedded into the general PFB of Hilbert Schmidt operators considered by Uhlmann. This allows us to construct a curvature on the constrained $U(n)$ EOPS bundle which is directly inherited from Uhlmann's general construction.
By general definition, a curvature $\mathcal F_{\mu\nu}$ is the geometric phase per area associated with the parallel transport around and infinitesimal parallelogram $\gamma_{\mu\nu}$ spanned by the vectors $\delta_\mu \hat e_\mu,\delta_\nu \hat e_\nu$ in momentum space (or more generally in parameter space). To obtain the curvature of the constrained EOPS bundle, we have to project Uhlmann's geometric phase $H_U^{\gamma_{\mu\nu}}$ (see Eq. (\ref{eqn:Uhlmannhol})) associated with the infinitesimal loop $\gamma_{\mu\nu}$ for $\hat \rho$ onto the $n$-dimensional subspace of interest by virtue of $\mathcal P$. We denote by $\hat H_U^{\gamma_{\mu\nu}}$ the representation matrix of the geometric phase in this subspace, i.e.,  $\left(\hat H_U^{\gamma_{\mu\nu}}\right)_{ij}=\langle \psi_i\rvert H_U^{\gamma_{\mu\nu}} \lvert \psi_j\rangle$. With these definitions, the curvature of the $U(n)$ EOPS bundle is given by
\begin{align}
\hat{\mathcal F}_{\mu\nu}=\lim_{\delta\rightarrow 0}\frac{1-\hat H_U^{\gamma_{\mu\nu}}}{\delta_\mu\delta_\nu},
\label{eqn:curvdef}
\end{align}
where $\delta \rightarrow 0$ is shorthand for $(\delta_\mu,\delta_\nu)\rightarrow (0,0)$. In particular, the gauge invariant Abelian curvature reads as
\begin{align}
\hat{\mathcal  F}_{\mu\nu}^A= \text {Tr} \hat{\mathcal F}_{\mu \nu}=-\lim_{\delta\rightarrow 0}\frac{\log\det(\hat H_U^{\gamma_{\mu\nu}})}{\delta_\mu\delta_\nu},
\end{align}
where $\log\det(\hat H_U^{\gamma_{\mu\nu}})$ is $i$ times the argument of the Abelian geometric phase factor.
In terms of this curvature, standard topological invariants for mixed states with spectral constraints can be defined.
For example, if the constrained rank $n$ density matrix $\hat \rho(k)$ is parameterized by a lattice momentum $k$ in a two-dimensional Brillouin zone (BZ), the first Chern number is given by
\begin{align}
\mathcal C=\int_{\text{BZ}}\frac{i\hat{\mathcal F}^A}{2\pi}=\int_{\text{BZ}}\text{d}^2k\,\frac{i\hat{\mathcal F}^A_{xy}}{2\pi}.
\label{eqn:Cherndef} 
\end{align}
The generalization to other invariants and the implementation of physical symmetries is straight forward and analogous to the pure state case. As a prominent example, we would like to mention the case of time reversal symmetry as a
protecting symmetry (see Section \ref{sec:gtot} for a discussion of the pure state analog). Instead of the projection onto the occupied bands, the density matrix $\rho(k)$ itself then obeys the symmetry $\mathcal T\rho(k)\mathcal T^{-1}=\rho(-k)$ with the anti-unitary time reversal operator $\mathcal T$. As in the pure state case, this symmetry carries over to arbitrary Uhlmann-holonomies and allows for the definition of topological invariants in analogy to Ref. \cite{Prodan2011}.

\section{Uhlmann phase winding numbers}
\label{sec:comments}
In two recent back-to-back publications \cite{Arovas2014, Delgado2014}, a complementary approach towards the definition of topological invariants for mixed states has been reported, in particular for Chern insulators in thermal states. Their approach is based on the so called Uhlmann phase $\phi_U^\gamma$ associated with a loop $\gamma$ in momentum space, defined as
\begin{align}
\text{e}^{i\phi_U^\gamma}&=\text{Tr}\left[w(0)^\dag w(T)\right]=\text{Tr}\left[\rho(0)U(T)U(0)^\dag\right]\nonumber\\&=\text{Tr}\left[\rho(0)H_U^\gamma\right],
\label{eqn:phiUhlmann}
\end{align}
where $\gamma$ is traversed between $t=0$ and $t=T$, $w(t)=\sqrt{\rho(t)}U(t)$ is a parallel transport in the sense of Uhlmann's connection (\ref{eqn:UhlmannCon}) and the Uhlmann holonomy $H_U^\gamma$ has been defined in Eq. (\ref{eqn:Uhlmannhol}). By construction, $\phi_U^\gamma$ bears some analogy with the Abelian Berry phase of a pure state defined as $\text{e}^{i\phi^\gamma}=\langle \psi(0)\vert \psi(T)\rangle$ (cf. Section \ref{sec:pureintro}). However, there is one crucial difference in the mathematical structure of $\phi^\gamma$ and $\phi_U^\gamma$: While $\text{e}^{i \phi^\gamma}$ is a $U(1)$-holonomy as is clear from the explicit representation in terms of an integral over a $U(1)$-gauge field (see Eq. (\ref{eqn:BerryPhase})), $\text{e}^{i\phi_U^\gamma}$ does in general {\emph{not}} have this property -- in contrast to the full non-Abelian Uhlmann holonomy $H_U^\gamma$. This observation is not just a minor technical point but can have drastic consequences. The key feature of a $U(1)$-holonomy is its additive group structure, i.e., $\phi^{\gamma_{12}}=\phi^{\gamma_{1}} + \phi^{\gamma_2}~(\text{mod} 2\pi)$, where $\gamma_{12}$ denotes the concatenation of the loops $\gamma_1$ and $\gamma_2$. It is this very property that allows us to go from an infinitesimal  geometric phase as measured locally by a curvature (see Eq. (\ref{eqn:curvdef})) to a quantized global topological invariant represented as an integral over the entire parameter space (see, e.g., the definition of the Chern number in Eq. (\ref{eqn:Cherndef})). Again making use of this additive group structure of the ordinary Berry phase, the Chern number $\mathcal C$ can be represented as the change of the Berry phase $\phi^{k_x}=\int_{S^1}\text{d}k_y\,\mathcal A_{y}(k_x,k_y)$ associated with a $k_y$-circle in the BZ at fixed $k_x$. This reformulation is achieved by dividing the BZ torus into $k_y$-rings $R_{k_x}$ of infinitesimal width $[k_x,k_x+dk_x]$ around $k_x$ and then using the Stokes theorem to transform the area integral of the curvature over these rings into a line integral of the connection $\mathcal A$ along their boundaries, i.e., $\int_{R_{k_x}}\text{d}^2k \mathcal F_{xy}=\int_{S^1}\text{d}k_y\left[\mathcal A_y(k_x+dk_x,k_y)-\mathcal A_y(k_x,k_y)\right]=\left(\frac{\partial \phi^{k_x}}{\partial k_x}\right)dk_x$. Explicitly, this procedure leads to the following well known representation of the Chern number
\begin{align}
\mathcal C = \frac{1}{2\pi}\int_{S^1}
\left(\frac{\partial \phi^{k_x}}{\partial k_x}\right)\, \text{d} k_x,
\label{eqn:Chernalt}
\end{align}  
where the circle $S^1$ is around the $k_x$ loop of the BZ. Equivalently, the role of $k_x$ and $k_y$ could be exchanged here, of course. This is again because of the holonomy group structure of the Berry phase which allows us to divide the BZ into stripes in an arbitrary direction.

The basic idea of Refs. \cite{Arovas2014,Delgado2014} is to take the right hand side of Eq. (\ref{eqn:Chernalt}) but to replace the ordinary Berry phase by the Uhlmann phase as defined in Eq. (\ref{eqn:phiUhlmann}), i.e.,
\begin{align}
\mathcal C_U = \frac{1}{2\pi}\int_{S^1}
\left(\frac{\partial \phi^{k_x}_U}{\partial k_x}\right)\, \text{d} k_x.
\label{eqn:ArovasDelgado}
\end{align}
We note that Ref. \cite{Arovas2014} actually discusses several constructions in terms of the spectrum of the ``holonomy matrix" $\rho(0)H_U^\gamma$. Since  the absence of a holonomy group structure which is at the heart of our present discussion also pertains to $\rho(0)H_U^\gamma$, this distinction is not of central importance here as will be addressed more explicitly in our example below.
As both $k_x$ in the BZ and the phase $\phi_U^{k_x}$ are defined on a circle, the mapping $k_x \mapsto \phi_U^{k_x}$  is characterized by an integer quantized winding number which is exactly measured by Eq. (\ref{eqn:ArovasDelgado}). 

However, since the Uhlmann phase does not have an additive group structure, $\mathcal C_U$ cannot be represented as the integral of a curvature over the 2D BZ. It is hence not immediately clear to what extent $\mathcal C_U$, technically being a 1D winding number, can be seen as a unique property of the 2D system under investigation. In particular, it is not clear that $\mathcal C_U$ as defined in Eq. (\ref{eqn:ArovasDelgado}) is equal to $\tilde{\mathcal C}_U$ obtained from $\mathcal C_U$ by exchanging $k_x$ and $k_y$ in all calculations, i.e.
\begin{align}
\tilde{\mathcal C}_U=\frac{1}{2\pi}\int_{S^1}
\left(\frac{\partial \phi^{k_y}_U}{\partial k_y}\right)\, \text{d} k_y.
\label{eqn:cutilde}
\end{align}

\begin{figure}[htp]
\centering
\includegraphics[width=0.45\columnwidth]{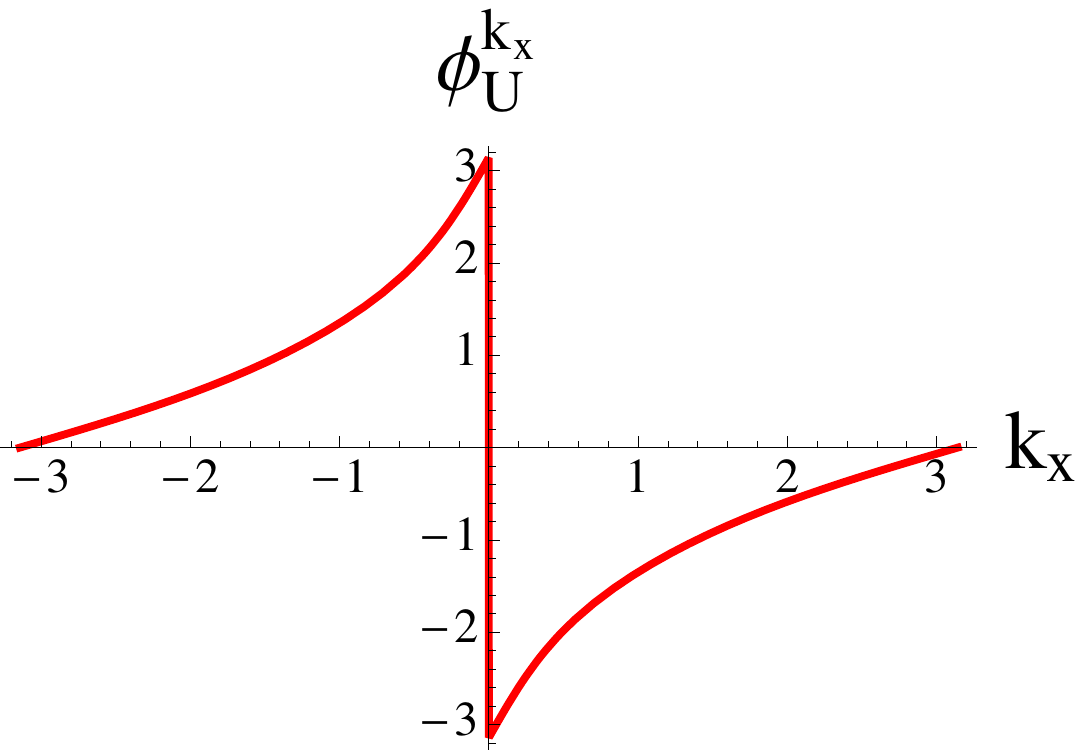}
\hspace{5mm}
\includegraphics[width=0.45\columnwidth]{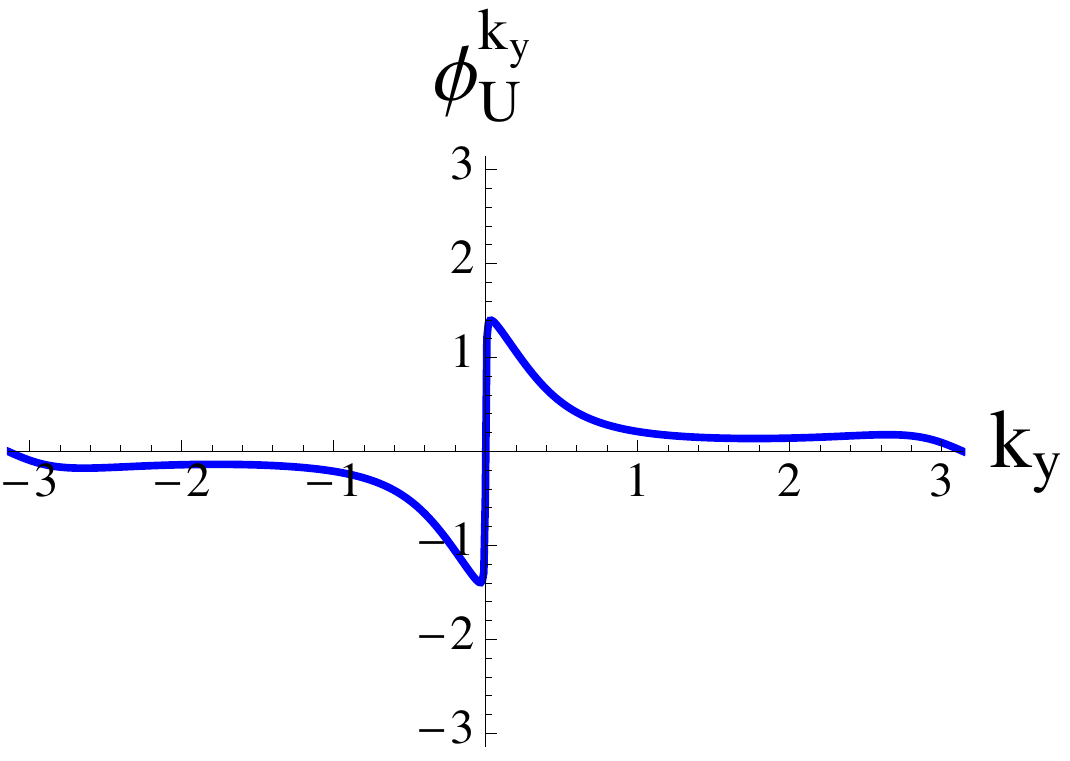}
\caption{\label{fig:xycomparison}(color online) Momentum dependence of the Uhlmann phase for a thermal state at $\beta=1.3$ for the model defined in Eq. (\ref{eqn:modelaniso}), numerically obtained by discretizing the paths in momentum space into $500$ equidistant points. Left panel: $\phi_U^{k_x}$ as a function of $k_x$. Right panel: $\phi_U^{k_y}$ as a function of $k_y$.} 
\end{figure}

Here, we demonstrate that indeed $\mathcal C_U\ne \tilde{\mathcal C}_U$ can occur for a Chern insulator the Hamiltonian of which is not invariant under the exchange of $k_x$ and $k_y$.
To this end, we consider a two-banded fermionic Chern insulator on a 2D square lattice with unit lattice constant defined by the Bloch Hamiltonian
\begin{align}
&H(k)=\vec d(k)\cdot \vec \sigma=\sum_{j=1}^3 d^j(k)\sigma_j,\nonumber\\
&d^1(k)=\sin(k_x),~d^2(k)=3\sin(k_y),\label{eqn:modelaniso}\\
&d^3(k)=1-\cos(k_x)-\cos(k_y)\nonumber,
\end{align}
where $\sigma_j$ are Pauli matrices. Note the anisotropy factor of $3$ in $d^2(k)$ which is crucial here. For the {\emph{two-banded}} model (\ref{eqn:modelaniso}), the Uhlmann connection and the Uhlmann phase can be readily calculated along the lines of Ref. \cite{Huebner1993}. Explicitly, the Uhlmann connection $\mathcal A_U$ in this special case reads as
\begin{align}
\mathcal A_{U,\mu} = -\left[(\partial_{k_\mu}\sqrt{\rho(k)}),\sqrt{\rho(k)}\right].
\label{eqn:UhlmannTwoband}
\end{align} 
At $T=0$, Eq. (\ref{eqn:UhlmannTwoband}) concurs with the Kato connection \cite{Kato1950} (see Eq. (\ref{eqn:Kato})) for the pure state case and the ordinary Chern number (\ref{eqn:Chernalt}) reads as
\begin{align*}
\mathcal C=\frac{1}{4\pi}\int_{\text{BZ}}\text{d}^2k\,\left (\hat d(k)\cdot [(\partial_{k_x} \hat d(k))\times(\partial_{k_y} \hat d(k))]\right)
\end{align*}
with $\hat d=\frac{\vec d}{\lvert \vec d\rvert}$. Explicit calculation yields $\mathcal C=-1$. Also, both $\mathcal C_U$ and $\tilde{\mathcal C}_U$ trivially concur with $\mathcal C$ since the Uhlmann phase reduces to the Berry phase for pure states such that Eq. (\ref{eqn:Chernalt}), Eq. (\ref{eqn:ArovasDelgado}), and Eq. (\ref{eqn:cutilde}) become equivalent.
At finite $T$, the system is in a thermal state defined by $\rho(k)=\frac{1}{Z}\text{e}^{-\beta H(k)}$ with $Z=\text{Tr}[\text{e}^{-\beta H(k)}]$. Using the simplified form (\ref{eqn:UhlmannTwoband}) of the Uhlmann connection, the direct calculation of Uhlmann holonomies (\ref{eqn:Uhlmannhol}) for arbitrary loops in momentum space is straightforward which in turn allows the direct calculation of $\mathcal C_U$ and $\tilde{\mathcal C}_U$ from Eq. (\ref{eqn:ArovasDelgado}) and Eq. (\ref{eqn:cutilde}), respectively. In the infinite temperature limit, this calculation becomes trivial as $\rho(k)=\mathbb 1$ independent of $k$ and hence $\mathcal A_U=0$ in Eq. (\ref{eqn:UhlmannTwoband}), resulting in $\mathcal C_U=\tilde{\mathcal C}_U=0$. Hence, the Uhlmann phase winding numbers have to jump from $-1$ to $0$ at some temperature. Numerically performing this calculation for finite temperature, we find that both $\mathcal C_U$ and $\tilde{\mathcal C}_U$ jump to zero at finite critical temperatures $\beta_c =\frac{1}{T_c}$ and $\tilde \beta_c =\frac{1}{\tilde T_c}$, respectively. Remarkably, for the present model, $\beta_c\ne \tilde \beta_c$, i.e., $\mathcal C_U$ and $\tilde{\mathcal C}_U$ do not concur at all temperatures. Numerically, we find $\beta_c=0.874$ and $\tilde \beta_c=1.32$. This discrepancy is illustrated In Fig. \ref{fig:xycomparison}, where we visualize the winding numbers $\mathcal C_U$ (left panel) and $\tilde{\mathcal C}_U$ (right panel) for $\beta=\frac{1}{T}=1.3$, i.e., $\beta_c<\beta<\tilde \beta_c$. Clearly, $\mathcal C_U=1$ since $\phi_U^{k_x}$ monotonously increases, interrupted by a jump from $\pi$ to $-\pi$ at $k=0$, by $2\pi$ as $k_x$ completes the loop from $k_x=-\pi$ to $k_x=\pi$. In contrast,  $\phi_U^{k_y}$ does not reach all values in $[-\pi,\pi]$ which implies $\tilde{ \mathcal C}_U=0$.  On a more detailed note, we would like to point out that, for the parameters chosen here, $\mathcal C_U$ is in the well defined regime in the sense of Ref. \cite{Arovas2014}. This is because the ``holonomy matrix" $\rho(k_x,k_{y0})H_U^{\gamma_{k_x,k_{y0}}}$ of a $k_y$-loop with footpoint $k_{y0}$ at fixed $k_x$ is gapped for all $k_x,k_{y0}$, i.e., the difference of the absolute values of its eigenvalues is finite and bounded from below by $\Delta= 0.268$. Comparing the construction in Refs. \cite{Arovas2014,Delgado2014} to our present analysis in Section \ref{sec:mixtop}, we would like to emphasize that the Chern number defined through Eq. (\ref{eqn:curvdef}) and Eq. (\ref{eqn:Cherndef}) does not exhibit similar finite temperature transitions as $\mathcal C_U$ and $\tilde{\mathcal C}_U$ but can, for the model (\ref{eqn:UhlmannTwoband}), be uniquely defined as $-1$ for the larger eigenvalue of the density matrix for all finite temperatures. At infinite temperature, in contrast, the density matrix does not obey any spectral constraints in the sense discussed in Section \ref{sec:speccon}, rendering all topological invariants trivial.\\

\section{Concluding remarks}
\label{sec:conclusion}
We have discussed how several assumptions regarding the spectrum of a family of density matrices can lead to a topologically non-trivial gauge structure.
In this framework topological invariants that are protected by these spectral assumptions have been defined for mixed states. Protected here means that topologically inequivalent mappings from a parameter space into the density matrices can be continuously deformed into each other only if the underlying spectral assumptions are violated. Identification of the parameter space with the Brillouin zone of a lattice translation invariant system provides one way of generalizing topological band structure invariants to the realm of mixed states. A non-trivial example where going beyond pure states is crucial to obtain a topologically non-trivial Chern insulator as a steady state in the framework of a non-equilibrium open quantum system dynamics has been reported in Ref. \cite{DissCI}. 
Additional physical symmetries refining this system of topological invariants can be considered in analogy to the pure state case.
The topological invariants defined here, being gauge invariant properties of the density matrix, are in principle experimentally accessible via state tomography.
However, their relation to natural observables such as response functions is not yet conclusively understood. First progress along these lines has been reported in Ref. \cite{Avron2012}. As for two-banded Chern insulators such as the toy model in Eq. (\ref{eqn:modelaniso}), any statistical mixture of bands with opposite Chern number will certainly cause deviations from the quantized Hall conductance. In other situations where the physical ramification of the topological invariant is related to a half-integer quantized polarization, a statistical mixture may still exhibit a quantization.\\

\section{Acknowledgements}
{Support from the ERC \je{Synergy Grant} UQUAM, and the START Grant No. Y 581-N16 is gratefully acknowledged.}

\bibliographystyle{apsrev}

\begin{thebibliography}{7}
\bibitem{note2}
The mathematical background of this statement is that for every smooth vector bundle, there is a ``complement" such that the direct sum of the two gives a trivial bundle over the same base manifold.
\bibitem{note1}
$R(t)$ may be thought of as external electric or magnetic fields for example. A finite spacing between the energy levels at all times is assumed to define the adiabatic time scale.
\end{thebibliography}

\end{document}